\documentclass[%
 reprint,
superscriptaddress,
 amsmath,amssymb,
 aps,
prb,
]{revtex4-1}

\usepackage[utf8] {inputenc}
\usepackage[english] {babel}
\usepackage{graphicx}
\usepackage{dcolumn}
\usepackage{mathtools}
\usepackage{float}
\usepackage{multirow}
\usepackage{amsmath}

\usepackage{lmodern}
\usepackage{textcomp}
\usepackage{t1enc}

\usepackage{graphicx}
\usepackage{setspace}

\usepackage[toc, page]{appendix}
\usepackage{standalone}

\usepackage{verbatim}

\usepackage{natbib}

\begin{document}


\title{Characterization and formation of NV centers in 3C, 4H and 6H SiC: an \textit{ab initio} study}

\author{A.\ Cs\'or\'e}
\affiliation{%
 Department of Atomic Physics, Budapest University of Technology and Economics, Budafoki út 8., H-1111, Budapest, Hungary}

\author{H.\ J.\ von Bardeleben}
\affiliation{%
Sorbonne Universit\'es, UPMC Universit\'e Paris 06, CNRS-UMR 7588, Institut des NanoSciences de Paris, 75005 Paris, France}

\author{J.\ L.\ Cantin}
\affiliation{%
Sorbonne Universit\'es, UPMC Universit\'e Paris 06, CNRS-UMR 7588, Institut des NanoSciences de Paris, 75005 Paris, France}

\author{A.\ Gali}
\affiliation{%
 Wigner Research Centre for Physics, Hungarian Academy of Sciences, PO. Box 49, Budapest H-1525, Hungary}
 \affiliation{%
 Department of Atomic Physics, Budapest University of Technology and Economics, Budafoki út 8., H-1111, Budapest, Hungary}
\email{gali.adam@wigner.mta.hu} 

\date{\today}

\begin{abstract}
Fluorescent paramagnetic defects in solids have become attractive systems for quantum information processing in the recent years. One of the leading contenders is the negatively charged nitrogen-vacancy defect in diamond with visible emission but alternative solution in technologically mature host is an immediate quest for many applications in this field. It has been recently found that various polytypes of silicon carbide (SiC), that are standard semiconductors with wafer scale technology, can host nitrogen-vacancy defect (NV) that could be an alternative qubit candidate with emission in the near infrared region. However, it is much less known about this defect than its counterpart in diamond. The inequivalent sites within a polytype and the polytype variations offer a family of NV defects. However, there is an insufficient knowledge on the magneto-optical properties of these configurations. Here we carry out density functional theory calculations, in order to characterize the numer
 ous forms of NV defects in the most common polytypes of SiC including 3C, 4H and 6H, and we also provide new experimental data in 4H SiC. Our calculations mediate the identification of individual NV qubits in SiC polytypes. In addition, we discuss the formation of NV defects in SiC with providing detailed ionization energies of NV defect in SiC which reveals the critical optical excitation energies for ionizing this qubits in SiC. Our calculations unravel the challenges to produce NV defects in SiC with a desirable spin bath.

\end{abstract}

\maketitle


\section{\label{sec1}Introduction}

In recent years spin carrying defects in solids have proved to be highly suitable for qubit \cite{quantumcomp, qubit} and nanoscale sensor applications \cite{sensor1}. So far the most investigated defect is the negatively charged nitrogen-vacancy defect (NV center) in diamond \cite{NVdiamond, NVdiamond1, nanosens} for which the afore-mentioned applications have been achieved. The exceptional properties  of the NV center in diamond is related to its optically polarizable spin triplet ($S = 1$) ground state ($^3A_2$), a spin dependent radiative recombination from the excited $^3E$ triplet state and a parallel operating spin selective non-radiative decay via intermediate singlet states ($^1A_1$, $^1E$). This mechanism results in the strong spinpolarization of the $^3A_2$ ground state with a predominant population of the $m_s = 0$ state. In addition, the spinstate of the $^3A_2$ state can be read out optically even at room temperature due to the long spin-coherence times\cite{NVdiamond, 
 divacexp} via the optically detected magnetic resonance (ODMR) effect.

Despite the unique properties of NV center in diamond, the material properties of diamond are not optimal and difficult to integrate into existing semiconductor device technology. In addition, the visible emission of NV center in diamond is not favorable for quantum communication where the fiber optics provides the most efficient transmission at near infrared (NIR) wavelengths.  Alternative qubits in technologically mature materials for various quantum technology application are much sought after. One of the most favorable candidates is silicon carbide with hosting divacancy defects that consist of adjacent carbon and silicon vacancies in the SiC lattice \cite{galipssb2011, quantumcomp, divacexp, Falk2013NatCom, arxivChristle}. This defect exhibits very similar properties to those of NV center in diamond, including the optical coherent control of this $S=1$ center \cite{quantumcomp}, and a relatively high contrast optical readout at resonant excitation \cite{arxivChristle}, but it emits in NIR region (around $1100$~nm of wavelength)  not far from the telecom wavelengths. NIR emission is also desirable for \textit{in vivo} fluorescent biosensor applications where fabrication of nanocrystalline SiC hosting NIR color centers were already suggested to this end \cite{NIR, somogyi_JPC}.  

Very recently, the equivalent of the NV center in diamond, the N$_\text{C}$V$_\text{Si}$ centers have been identified in different (3C,4H,6H) polytypes of SiC \cite{4Hexp, NVSiCPL, NVSiCour}. In the negative charge state they are  spin S=1 centers with optical properties shifted to the  NIR region (around $1200$~nm of wavelength) almost compatible with the transmission wavelength of optical fibers.  However, some of their optical properties and excited state configurations have not yet been fully resolved either in experiments \cite{4Hexp,NVSiCour} or in theory \cite{quantumcomp, NVformation, NVgordon_PRB, NVSiCour} and thus required further investigations. Since qubits are individual quantum objects, thorough characterization of the individual NV defects in SiC is required to optimize the conditions of qubit operations. To this end, we carried out density functional theory (DFT) calculations of nitrogen-vacancy defects in 3C, 4H and 6H SiC. We provide a detailed results concerning their electronic structure, magneto-optical parameters, ionization energies and formation energies. We also discuss the possible formation processes of NV defects in SiC and briefly compare the formation and ionization energies of NV center and divacancy defects in SiC. 

In the followings, we describe the computational methods in Section~\ref{sec:methods}. Our results are presented in detail in Section~\ref{sec:results} where we compare them to the experimental data if available. Here, we show new electron paramagnetic resonance (EPR) spectra for basal NV centers in 4H SiC. We discuss the photo-ionization and the formation of NV center in SiC in Section~\ref{sec:discussion}. Finally, we conclude our work in Section~\ref{sec:summary}.

\section{\label{sec2}Defect structure}

SiC exhibits various crystal structures, called polytypes, with 4H, 6H and 3C the most advanced from a material point of view. We have therefore calculated the properties of NV center in these technologically important polytypes. We found by \emph{ab initio} calculations that formation energy of N$_{\text{C}}$V$_{\text{Si}}$ defect is lower by about 2~eV than that of N$_{\text{Si}}$V$_{\text{C}}$ defect in each considered polytype in the neutral charged state implying that nitrogen preferentially substitutes carbon (N$_{\text{C}}$) in the SiC lattice, adjacent to a neighbor Si-vacancy (V$_{\text{Si}}$). We show the results on N$_{\text{C}}$V$_{\text{Si}}$ in detail that we also call NV center in the context. 

\begin{figure}
\includegraphics[width=0.5\textwidth]{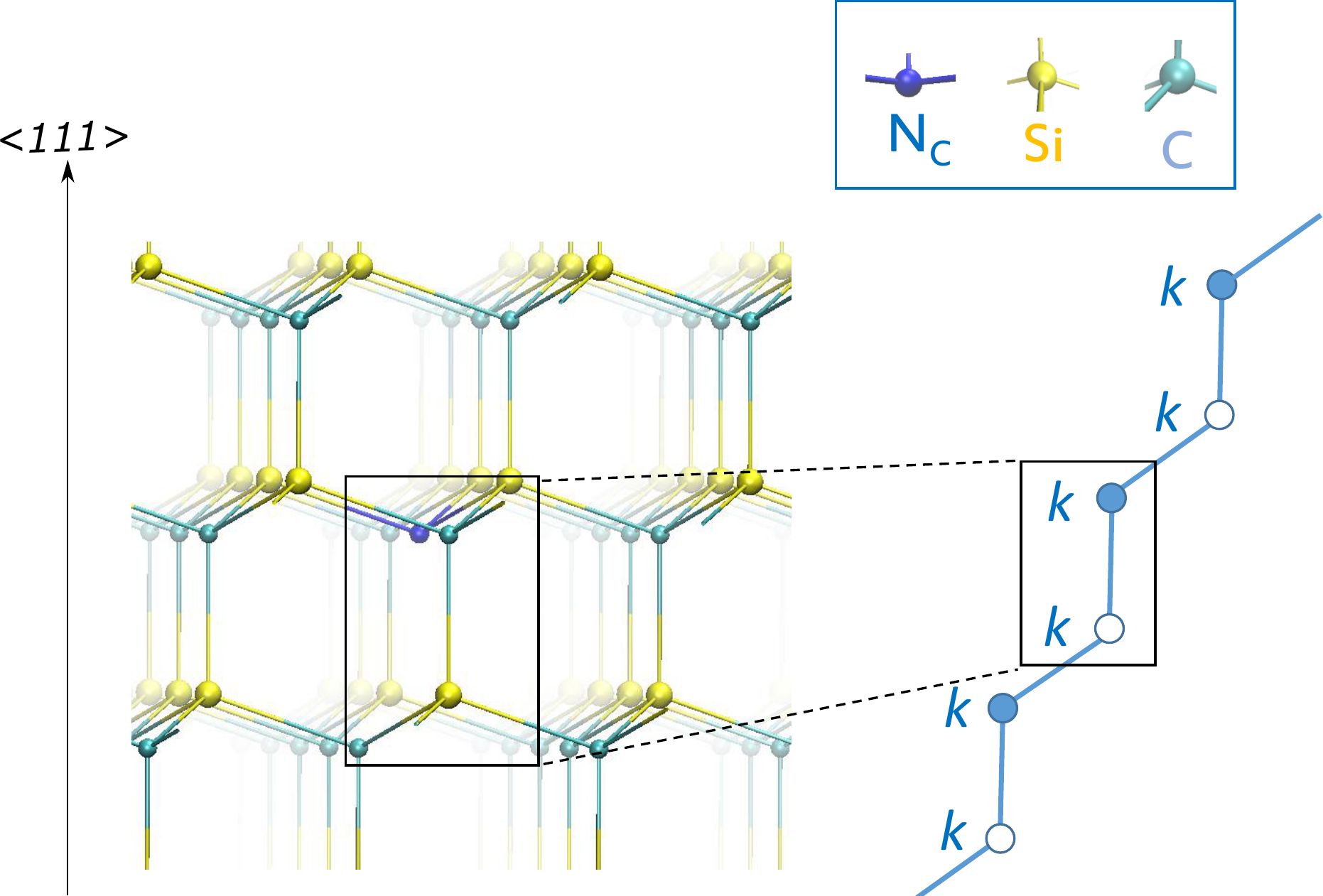}
\caption{Single configuration ($kk$) of N$_{\text{C}}$V$_{\text{Si}}$ defect  in 3C SiC}
\label{defstruct3c}
\end{figure}

\begin{figure}
\includegraphics[width=0.5\textwidth]{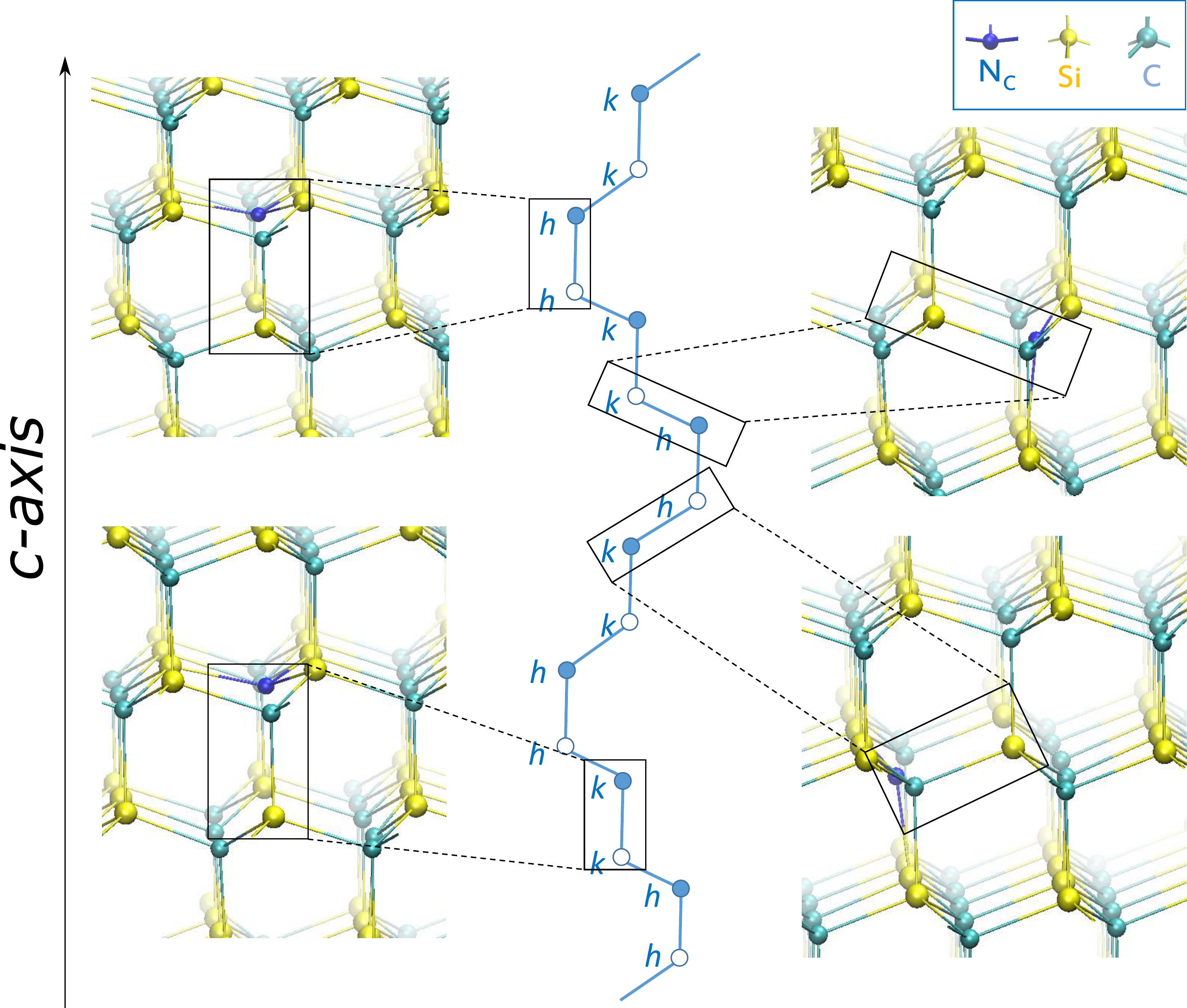}
\caption{Possible configurations of N$_{\text{C}}$V$_{\text{Si}}$ defect in 4H SiC. The $hh$ and $kk$ configurations exhibiting C$_{\text{3v}}$ symmetry are called \textit{on-axis} configurations, since the axis of defects is parallel to the c-axis, while $kh$ and $hk$ configurations are \textit{off-axis} geometries with C$_{\text{1h}}$ symmetry.}
\label{defstruct4h}
\end{figure}

\begin{figure}
\includegraphics[width=0.5\textwidth]{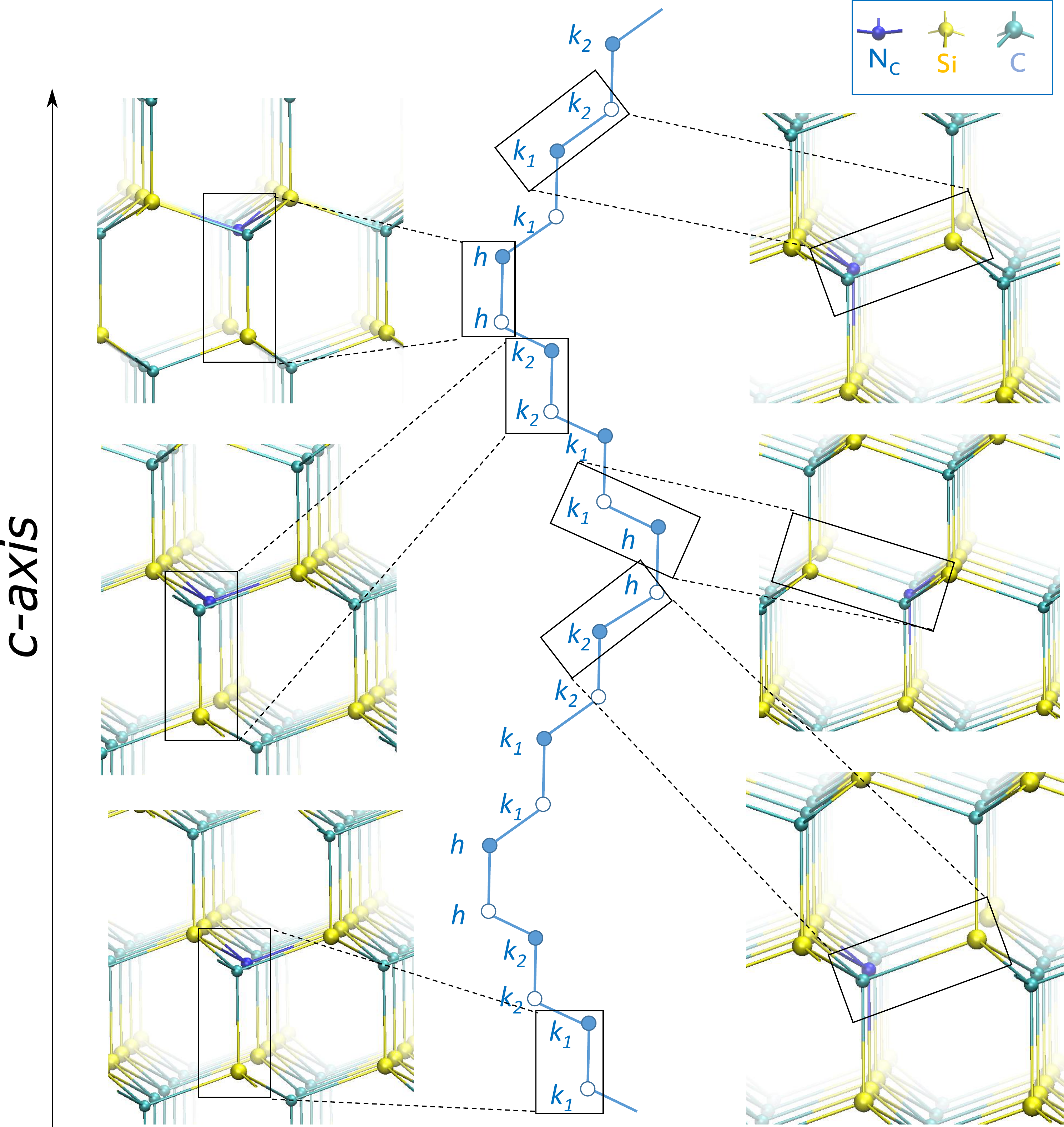}
\caption{Possible configurations of N$_{\text{C}}$V$_{\text{Si}}$ defect structure in 6H SiC. Three on-axis ($hh$, $k_1k_1$, $k_2k_2$) and three off-axis or basal configurations ($hk_1$, $k_1k_2$, $k_2h$) can be formed.}
\label{defstruct6h}
\end{figure}

Due to the special arrangement of the atoms in hexagonal polytypes, different varieties of NV center exist, depending on the lattice site of the N-atom and the adjacent Si-vacancy. In short, N$_{\text{C}}$V$_{\text{Si}}$ defect forms one configuration in 3C ($kk$) (cf. Fig. \ref{defstruct3c}), four configurations in 4H ($hh$, $kk$, $hk$, $kh$) (cf. Fig. \ref{defstruct4h}), and six configurations in 6H ($hh$, $k_1k_1$, $k_2k_2$, $hk_1$, $k_1k_2$, $k_2h$) (cf. Fig. \ref{defstruct6h}) polytype, where $h$ and $k_{\{1,2\}}$ labels the (quasi)cubic and hexagonal lattice sites in each polytype, respectively.  

\section{\label{sec:methods}Methodology}
\subsection{\label{subsec:comp}Computational approach}

Our calculations were carried out by means of HSE06 range-separated hybrid functional developed by Heyd, Scuseria and Ernzerhof \cite{HSE06}. In order to characterize ground and excited state zero-filed splitting (ZFS) arising from electron spin - electron spin dipole-diple interaction we calculated the \textit{D} and \textit{E} parameters employing the Perdew-Burke-Ernzerhof (PBE) \cite{PBE} functional as implemented by Iv\'ady \textit{et al.} \cite{ZFScalc}. In the excited state of on-axis defect configurations exhibits dynamic Jahn-Teller (JT) distortion due to an effective electron-phonon coupling, thus ZFS constants calculated under C$_\text{1h}$ symmetry were averaged to show a dynamic C$_\text{3v}$ symmetry. For basal configurations the natural symmetry is \emph{per se} C$_\text{1h}$ symmetry, nevertheless, similar electron-phonon coupling can occur in the excited state as that for the axial configurations. Therefore, we carried out a motional averaging procedure about the N-V axis for the basal NV configurations. In the calculation of zero-phonon lines (ZPL) we used the lowest total energy in the excited state corresponding to the JT geometry.

For atomistic modeling of defect structures, we applied 512-atom supercell for 3C, 576-atom supercell for 4H and 432-atom supercell for 6H polytype. For sampling the Brillouin-zone we employed $\Gamma$-point only for 3C and 4H polytypes, while $\Gamma$-centered $2\times2\times2$  Monkhorst-Pack \textit{k}-point mesh was used for 6H SiC. In addition, to reach sufficient accuracy for ZPL values $2\times2\times2$ \textit{k}-point mesh was employed in each case. Plane wave expansion of Kohn-Sham wavefunctions with a cutoff of 420~eV was applied. Relaxed geometries were achieved by minimizing the total energy with respect to the normal coordinates of the lattice using the force threshold of 0.01 eV/$\AA$. Core-electrons were treated by projector-augmented wave (PAW) \cite{PAW} potentials as implemented in VASP code \cite{VASP}. In the case of charged defects Freysoldt correction \cite{Freysoldt} in total energy was applied. The hyperfine couplings were calculated by HSE06 DFT calculations with taking into account the spin polarization of the core electrons \cite{hyperfine}.

\subsection{\label{subsec:form}Formation energies and charge transition levels}

Concentration of point defects in thermal equilibrium can be predicted via the formation energies. In addition, determining the adiabatic charge transition levels, i.e., ionization energies is also crucial to study the stability window of a given charge state of the defect that is applied as a qubit. We calculated the formation energies as \cite{ZhangPRL1991,aradimu}   
\begin{equation}
\begin{aligned}
E_{\text{form}}^q = E_{\text{tot}}^q - n_{\text{SiC}}\mu_{\text{SiC}} - \frac{\mu_{\text{Si}} - \mu_{\text{C}} - \delta\mu}{2}(n_\text{Si} - n_\text{C})\\
 - n_{\text{N}}\mu_{\text{N}} + q(E_{\text{f}}+E_\text{VBM}) + \Delta V(q),
\label{formeq}
\end{aligned}
\end{equation}
where $E_{\text{tot}}^q$ is the total energy of the defective system, $\mu_{\text{Si}}, \mu_{\text{C}}, \mu_{\text{N}}$ are the chemical potentials of Si atom in bulk Si, C atom in diamond and the N atom, respectively, $q$ is the charge state, $E_{\text{f}}$ is the Fermi-level, $E_\text{VBM}$ represents the valence band edge and $\Delta V(q)$ stands for the Freysoldt charge correction term \cite{Freysoldt}. If $\delta\mu$ is chosen to be the heat of formation of SiC, $ \mu_\text{SiC} - (\mu_\text{Si} +~\mu_\text{C})$, then this provides the formation energy of defects under stoichiometric conditions.  In the actual VASP parameters and implementation we obtained $\mu_\text{SiC}$ = -17.47~eV, $\mu_\text{Si}$ = -6.43~eV, $\mu_\text{C}$ = -10.55~eV and $\delta\mu$ = -0.49~eV by HSE06.  For determining $\mu_{\text{N}}$ the hexagonal $\beta$-Si$_3$N$_4$ was chosen as the most stable form of Si$_3$N$_4$ that can be considered as the solubility-limiting phase in SiC. We obtained $\mu_{\text{
 N}}$ = -12.22~eV upon these condition. The adiabatic charge transition levels can be derived from Eq.~\ref{formeq} as follows,
\begin{equation}
E_{q+1/q} = E_{\text{tot}}^q - E_{\text{tot}}^{q+1} + \Delta V(q) - \Delta V(q+1).
\label{eq:CHLs}
\end{equation}

Binding energy ($E_\text{binding}$) of defects A and B forming the complex AB can be defined as
\begin{equation}
E_\text{binding}(E_\text{f}) = E_\text{form}^{\text{A}}(E_\text{f}) + E_\text{form}^{\text{B}}(E_\text{f}) - E_\text{form}^{\text{AB}}(E_\text{f}).
\label{eq:binde}
\end{equation}
According to this definition $E_\text{binding}>0$ implies that the formation of AB complex is favorable.

\subsection{\label{subsec:conc}Calculation of defect concentrations}

Defects may be introduced during growth of the crystal. High quality silicon carbide is typically grown via chemical vapor deposition process that may be considered as a quasi-equilibrium process. The concentration of defects can be then estimated in thermal equilibrium at the growth temperatures. The concentration of a defect D$^q$ with charge state $q$ is the sum of the concentration of individual configurations D$_i^q$, i.e. symmetry inequivalent forms of D$^q$. This can be calculated by multiplying the number of possible defect sites per cm$^3$ ($N_\text{sites}^{\text{D}}$) by the statistical weight $\omega_{D^q}$ coming from the spin multiplicity and the Boltzmann-factor
\begin{equation}
[\text{D}^{q}(E_\text{f})] = \sum_\text{i}[\text{D}_i^q(E_\text{f})] = \frac{1}{Z}N_\text{sites}^{\text{D}}\omega_{D^q}\mathbf{\sum_i}\text{exp}\bigg(-\frac{E_\text{form}^{\text{D}_i^{q}}(E_\text{f})}{k_\text{B}T}\bigg),
\label{eq:conceq}
\end{equation}
where 
\begin{equation}
Z=1+\sum_{i,q}\text{exp}\bigg(-\frac{E_\text{form}^{\text{D}_i^{q}}(E_\text{f})}{k_\text{B}T}\bigg)
\label{eq:zeq}
\end{equation}
is the grand canonical partition function and the summation goes over all the individual configurations ($i$) and charge states ($q$) of defect D. In Eqs.~\ref{eq:conceq} and \ref{eq:zeq} $k_\text{B}$ is the Boltzmann constant and $T$ is the temperature. For calculating $E_\text{form}^{\text{D}_i^{q}}(E_\text{f})$ Fermi-energy has to be determined (Eq.~\ref{formeq}) which can be performed via solving the neutrality equation which reads as 
\begin{equation}
\begin{aligned}
N_c(T)\text{exp}\bigg(-\frac{E_\text{CBM}-E_\text{f}}{k_\text{B}T}\bigg) + \sum_{\substack{\text{D}\\
q<0}} |q_\text{D}|[\text{D}^{q}(E_\text{f})] = \\
 N_v(T)\text{exp}\bigg(-\frac{E_\text{f}-E_\text{VBM}}{k_\text{B}T}\bigg) + \sum_{\substack{\text{D}\\
q>0}} |q_\text{D}|[\text{D}^{q}(E_\text{f})],
\label{eq:neutraleq}
\end{aligned}
\end{equation}
where
\begin{equation}
N_c(T) = 2 \bigg(\frac{2m_\text{e}^*\pi k_\text{B}T}{h^2}\bigg)^{3/2}
\label{eq:neutraleqnc}
\end{equation}
and
\begin{equation}
N_v(T) = 2 \bigg(\frac{2m_\text{h}^*\pi k_\text{B}T}{h^2}\bigg)^{3/2}
\label{eq:neutraleqnv}
\end{equation}
are the effective densities of states of electrons in the conduction band edge and holes in the valence band edge, respectively, and $E_\text{CBM}, E_\text{VBM}$ labels the conduction and valence band edges, respectively. We applied the parameters of 4H SiC as we calculated the concentration of defects in this polytype Accordingly, for the effective masses in Eqs.~\ref{eq:neutraleqnc} and \ref{eq:neutraleqnv} we used $m_\text{h}^*$ = 1.26$m_\text{e}^0$ and $m_\text{e}^*$ = 0.39$m_\text{e}^0$, where $m_\text{e}^0$ is the electron rest mass. We note that the shallow substitutional nitrogen donors in SiC will be explicitly treated in Eq.~\ref{eq:conceq}.

In order to obtain Fermi energy the series of equations (Eqs.~\ref{eq:conceq} and \ref{eq:neutraleq}) has to be solved self-consistently. For the calculation thermal equilibrium and stoichiometric ratio of C and Si atoms were assumed. We considered infinite bulk material, thus, the effect of band bending near the surface and other kinetic effects were neglected. Bulk growth of SiC is usually carried out at temperatures between $1600 ^\circ$C and $2000 ^\circ$C \cite{temp, KordinaPSS1997}. Employing this technique different charge states of substitutional nitrogen (N$_\text{C}$), vacancies (V$_\text{C}$, V$_\text{Si}$), divacancies (V$_\text{C}$V$_\text{Si}$), carbon antisite vacancy pairs (CAV) and nitrogen-vacancy complexes ((N$_\text{C}$)$_k$V$_\text{Si}$, k=1,2,3,4) may be formed. For nitrogen-vacancy complexes we considered the case of $k$=1 (N$_\text{C}$V$_\text{Si}$) and the electrically inactive $k$=4 ((N$_\text{C}$)$_4$V$_\text{Si}$ defect) as the latter has an extremely low formation energy \cite{bockstedte1}. In this spirit, we calculated the concentrations of the these defects at temperatures of $1600 ^\circ$C, $1700 ^\circ$C, $1800 ^\circ$C, $1900 ^\circ$C and $2000 ^\circ$C considering all relevant charge states. To this end, we calculated the formation energies of all the defects in all configurations including V$_\text{Si}$, V$_\text{C}$ and CAV \cite{RaulsPSSB2011,singlephotonsourceNatMat,szaszCAV}.

\section{\label{sec:results}Results}

By using HSE06 functional we could reproduce the experimental band gap within 0.1~eV for 3C, 4H and 6H polytypes. This limits the accuracy of our method in the prediction of charge transition levels.

\subsection{\label{subsec32}Electronic structure}
 
\begin{figure} [h!]
\centering
\includegraphics[width=0.4\textwidth]{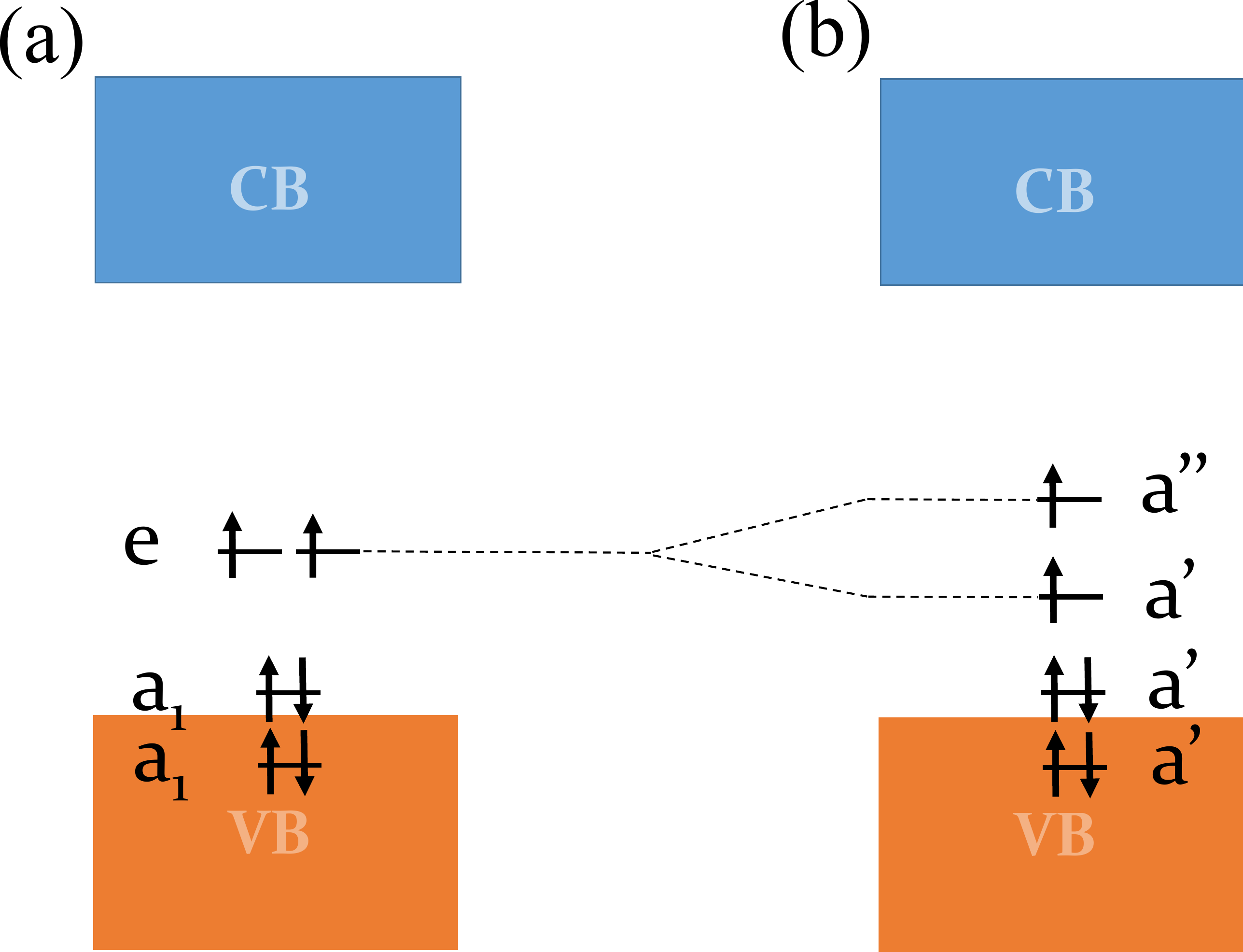}
\caption{\small Scheme of ground state electronic structure of (a) on-axis  and (b) off-axis NV center configurations exhibiting C$_\text{3v}$ and C$_\text{1h}$ symmetry, respectively.}
\vspace{- 20 pt}
\label{estruct}
\end{figure}

Group theory analysis on the defect provides intriguing insights its electronic structure. Accordingly, we found that on-axis configurations of NV center with C$_{\text{3v}}$ symmetry introducing two $a_1$ levels and a degenerate $e$ level. Off-axis configurations exhibit C$_\text{1h}$ symmetry, and the degenerate $e$ level splits to an $a'$ and an $a''$ state while the $a_1$ states transform into $a'$ states. Our calculations revealed that one of the $a_1$ levels falls in the valence band, whereas the other is lying in the fundamental band gap and they both are fully-occupied. In the single negative charged state, the degenerate $e$ level is introduced in the band gap occupied by two electrons with parallel spins providing $S = 1$ triplet spin state.  In summary, the one-electron structure of the ground state is $a_1(2)a_1(2)e(2)$ for on-axis and $a'(2)a'(2)a'(1)a''(1)$ for off-axis configurations in singly negative charge state (cf.\ Fig.~\ref{estruct}). Further application of grou
 p theory implies that $^3A_2$, $^1E$ and $^1A_1$ multiplets can be formed by the $a_1(2)a_1(2)e(2)$ electron configurations for the on-axis defects. The $^3E$ bright triplet excited are realized by $a_1(2)a_1(1)e(3)$ electron configuration, and an $^1E$ multiplet occurs too for on-axis defects. The $^3E$ excited state is a dynamic Jahn-Teller system. We calculated this excited state by $\Delta$SCF method with allowing $C_{1h}$ symmetry distortion but the zero-field constant was calculated in the dynamic average of $C_{3v}$ symmetry. For the off-axis configurations, the electronic configurations and states are similar to those of on-axis configurations, just the degenerates states are split due to the $C_{1h}$ symmetry crystal field.

\subsection{Formation energies and charge state stability of NV centers in SiC}
\label{ssec:form}

\begin{figure}
\centering
\includegraphics[width=0.5\textwidth]{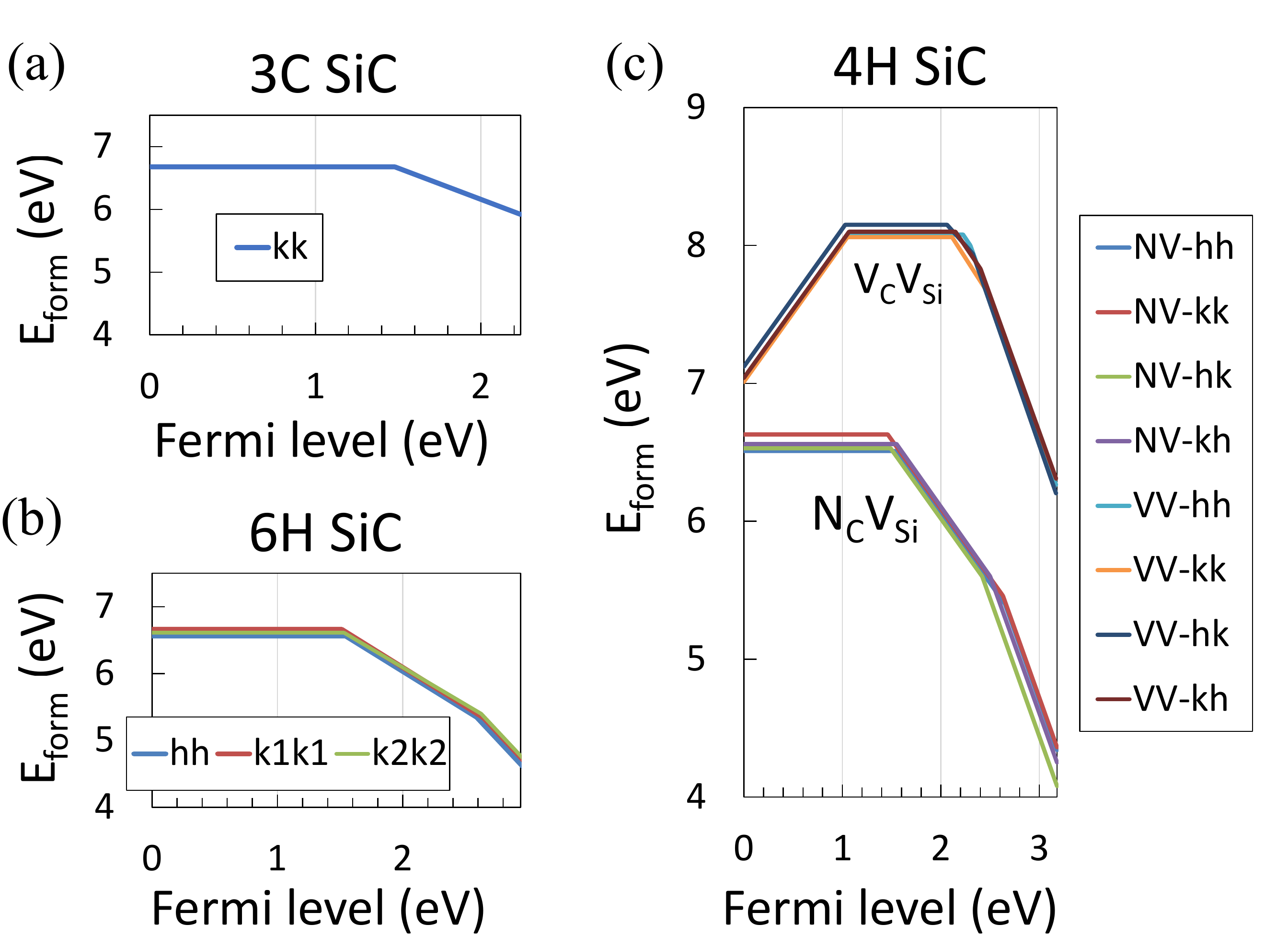}
\caption{Formation energies of the NV defects as a function of the position of the Fermi-level in (a) 3C, (b) 6H and (c) 4H polytypes. In 4H SiC we plot the formation energy of V$_{\text{C}}$V$_{\text{Si}}$ divacancy.}
\label{CTLfig}
\end{figure}

We plot the formation energies of the NV defect in the considered polytypes of SiC in Fig.~\ref{CTLfig}. We find that the neutral and negative charged states are stable as function of the position of the Fermi-level in all the polytypes, whereas the double negative charged state exists in hexagonal polytypes. The negatively charged NV defect, i.e., NV center can occur in moderately or highly n-type 3C SiC, whereas NV center is stable in non-doped or moderately n-type doped hexagonal SiC. In highly n-type doped hexagonal SiC, the NV defect becomes double negatively charged. This result implies that different doping strategies should be applied to stabilize the single negative charge state of the NV defect in cubic and hexagonal polytypes.

We find that the formation energies of N$_{\text{C}}$V$_{\text{Si}}$ defect configurations in 4H and 6H polytypes vary slightly, i.e., the values agree within $\sim0.1$~eV, whereas the maximum difference between the corresponding charge transition levels does not exceed $\sim0.2$~eV (see Table~\ref{CTLtab}). In Fig.~\ref{CTLfig}(c) formation energies of V$_{\text{C}}$V$_{\text{Si}}$ configurations in 4H polytype are also showed for discussion. 
\begin{table}[t]
\begin{ruledtabular}
\caption{Charge transition levels of  N$_{\text{C}}$V$_{\text{Si}}$ defects referenced to the valence band maximum ($E_{\text{VBM}}$).}
\label{CTLtab}
\begin{tabular}{@{}cccc@{}}
\multicolumn{1}{c}{Polytype}
& \multicolumn{1}{l} {conf.}
& \multicolumn{1}{c} {$E_{0/-}$ (eV)}
& \multicolumn{1}{c} {$E_{-/2-}$ (eV)}\\
\hline
3C & $kk$ & 1.48 &  \\
\hline
\multirow{4}{*} {4H} & $hh$ & 1.54 & 2.65 \\
& $kk$ & 1.46 & 2.63 \\
& $hk$ & 1.49 & 2.42 \\
& $kh$ & 1.55 & 2.50 \\
\hline
\multirow{3}{*} {6H} & $hh$ & 1.54 & 2.59 \\
& $k_1k_1$ & 1.51 & 2.64 \\
& $k_2k_2$ & 1.53 & 2.62
 \\
\end{tabular}
\end{ruledtabular}
\end{table}

\subsection{\label{ssec:DZPL}Magneto-optical properties of NV center in SiC}

The NV centers in all three polytypes have the same electronic structure with a $^3A_2$ triplet groundstate and a $^3E$ excited state. The paramagnetic groundstate make them suitable to EPR spectroscopy, which has been applied successfully for their assessment \cite{4Hexp, NVSiCPL, NVSiCour}.  Our calculations show in accordance with the experimental results that each type of NV center is characterized by specific zero-field-splitting (ZFS) parameters, hyperfine interactions (HF) and optical properties as shown in Tables~\ref{tab:hypertable}-\ref{tab:ZFStable6H}.
\begin{figure}
\includegraphics[width=0.4\textwidth]{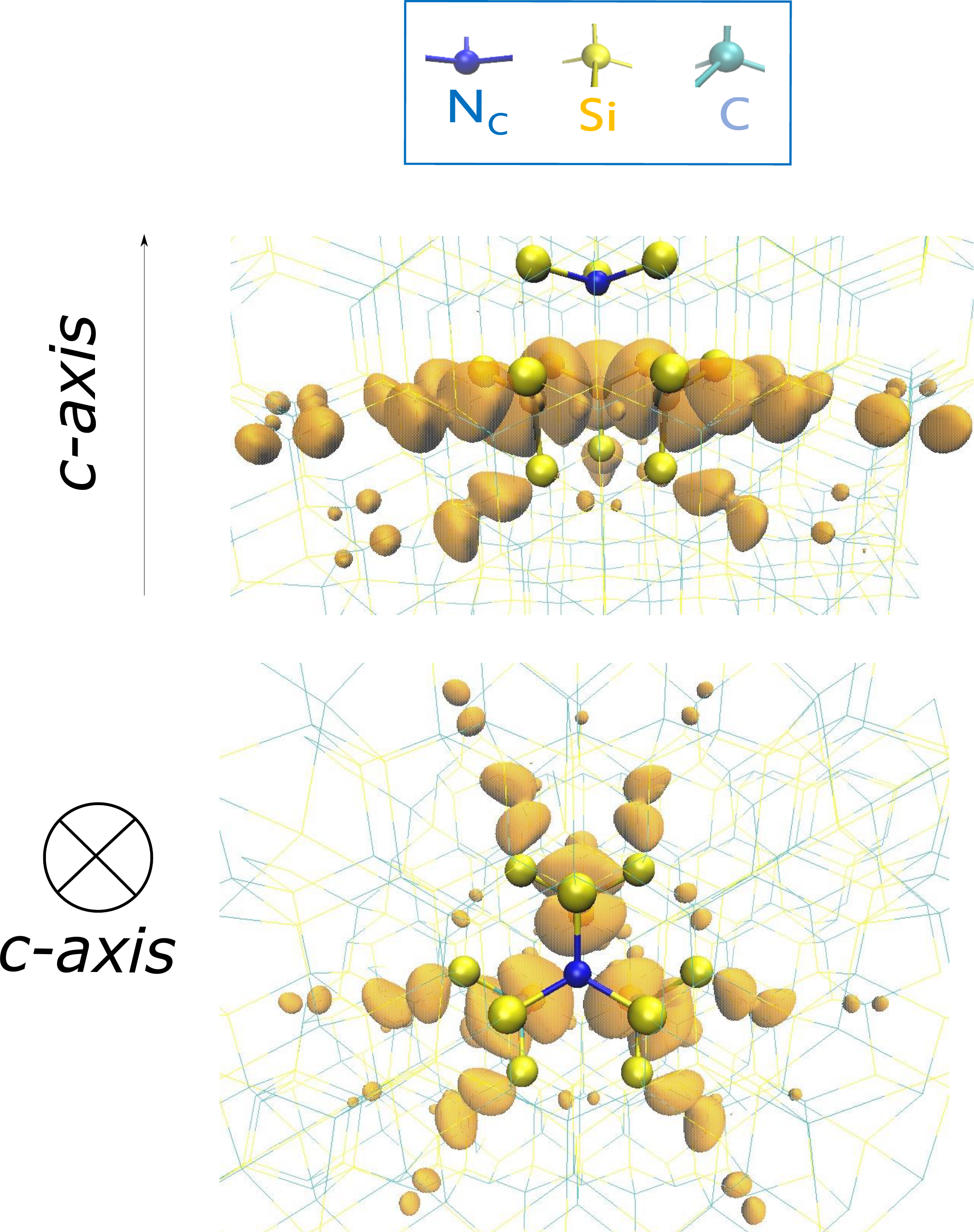}
\caption{Isosurface of the spin density localized on the three C-atoms near the C-vacancy in the triplet ground state of NV center in SiC. The supercell structure is shown in perspective view where the lattice is depicted as a wire except for the atoms in the core of the defect that are represented by balls. The corresponding atom types are labeled.}
\label{fig:spindens}
\end{figure}

We first discuss the spin density distributions. In the ground state, the major spin density is localized on the three neighbor C atoms of the Si-vacancy (cf.\ Fig.~\ref{fig:spindens}) providing a strong HF interaction with the $^{13}$C nuclear spins. The HF interactions with the $^{14}$N neighbor and its adjacent three $^{29}$Si nuclei are small ($\approx 1$~MHz). Nevertheless due to the small EPR linewidth of $0.02$~mT they are already resolved and represent the fingerprint of NV centers which distinguish them from other defects in SiC. The Si-atoms near the nearest neighbor C-atoms have stronger HF constants of $\approx 10$~MHz also resolved in the EPR spectra. The small value of the $^{14}$N HF constants is due to the only indirect HF interaction; the calculational accuracy for these values is smaller than for the HF with the $^{13}$C and $^{29}$Si neighbors with direct HF interaction and estimated to 10\% according to our tests on related defects \cite{hyperfine}. Nevertheless, 
 the calculated values are in good agreement with the experimental findings, as shown in Table~\ref{tab:hypertable}.
\begin{table*}
\begin{ruledtabular}
\caption{Calculated hyperfine constants ($A_{ii}; i=x,y,z$) in the ground state of on-axis NV center configurations. Experimental data\cite{NVSiCour} available for $^{14}\text{N}$ isotope are also listed.}
\label{tab:hypertable}
\begin{tabular}{@{}cccccccc@{}}
\multirow{2}{*}{Polytype}
& \multirow{2}{*} {conf.}
& \multicolumn{2}{c} {$1\times^{14}\text{N}$}
& \multicolumn{2}{c} {$3\times^{29}\text{Si}$ + $6\times^{29}\text{Si}$}
& \multicolumn{1}{c} {$1\times^{13}\text{C}$ }\\
& & $A_\text{exp}^\text{iso}$(MHz) & $A_{xx}, A_{yy}, A_{zz}$ (MHz) & $A_{xx}, A_{yy}, A_{zz}$ (MHz) & $A_{xx}, A_{yy}, A_{zz}$ (MHz) & $A_{xx}, A_{yy}, A_{zz}$ (MHz) \\ 
\hline
3C & $kk$ & 1.26 & -1.67, -1.67, -1.72 & 11.93, 11.91, 12.31 & 9.64, 8.47, 10.48 & 48.23, 47.49, 119.38  \\
\hline
\multirow{2}{*} {4H} & $hh$ & 1.23 & -1.57, -1.57, -1.64 & 11.80, 11.72, 12.14 & 9.95, 8.77, 10.79 & 45.49, 44.86, 116.78 \\
& $kk$  & 1.12 & -1.71, -1.69, -1.71 & 12.77, 12.39, 13.26 & 10.56, 9.60, 11.32 & 42.23, 41.48, 112.79\\
\hline
\multirow{3}{*} {6H} & $hh$ & 1.32 & -1.61, -1.61, -1.80 & 11.95, 11.84, 12.30 & 9.81, 8.74, 10.69 & 42.40, 41.81, 113.84\\
& $k_1k_1$ & 1.21 & -1.76, -1.75, -1.76 & 12.94, 12.63, 13.48 & 10.93, 10.17, 11.74 & 37.00, 36.33, 108.14 \\
& $k_2k_2$ & 1.26 & -1.73, -1.73, -1.76 & 12.05, 12.00, 12.48  & 10.51, 9.51, 11.39 &  42.44, 41.78, 114.36 \\
\end{tabular}
\end{ruledtabular}
\end{table*}

We also determined the ZFS parameters ($D$, $E$) for NV centers by assuming electron spin - electron spin dipolar interaction. We calculated these parameters in the  $^3A_2$ ground state for all the considered polytypes (Tables~\ref{tab:ZFStable3C4Ha}, \ref{tab:ZFStable6H}). In the point dipole spin - point dipole spin approximation, the $D$ varies as $D\sim\frac{1}{r^3}$ with $r$ being the distance between the dipoles. In the ground state, the spin density is localized on the C-atoms near the Si-vacancy, so the distance between these C-atoms gives the trend for the variation of the $D$-constants (see Tables~\ref{tab:ZFStable3C4Ha}, \ref{tab:ZFStable6H}). 

\begin{figure}
\includegraphics[width=\columnwidth]{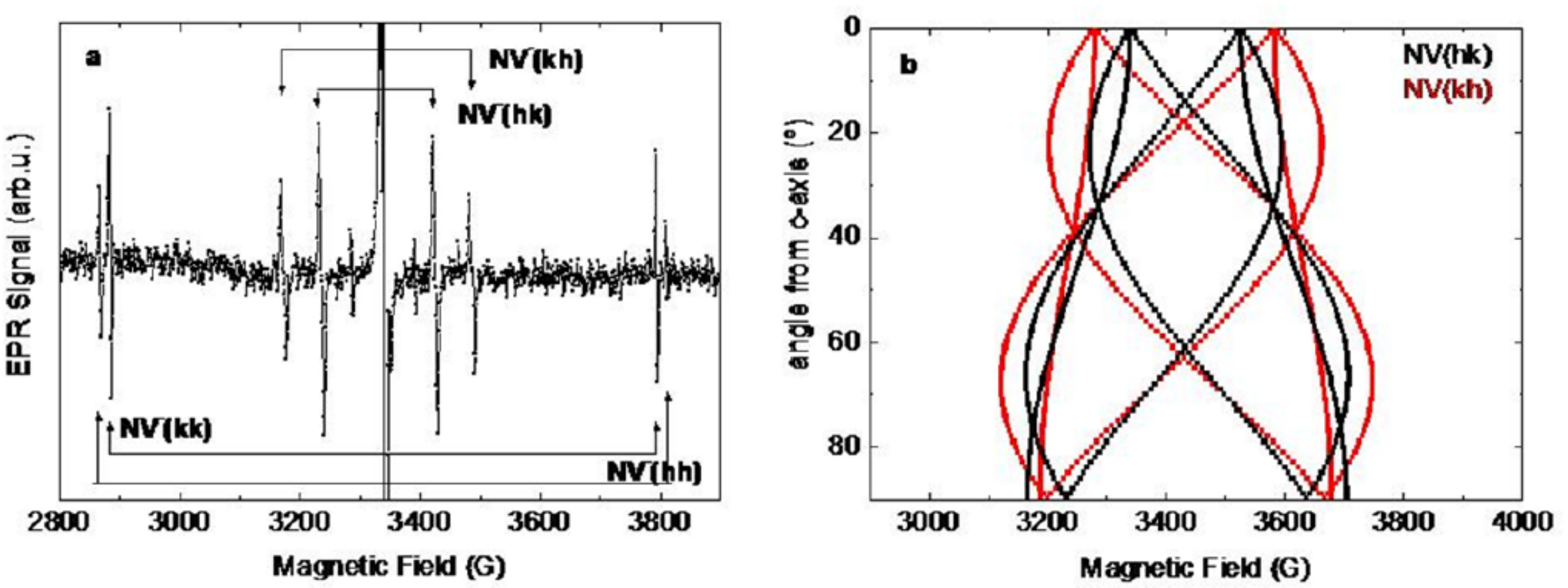}
\caption{(a) EPR spectrum of the NV center in 4H-SiC displaying the two axial and two basal related EPR spectra for $\textbf{B}\| c$ where $\textbf{B}$ is the applied external magnetic field and $c$ is the c-axis of 4H SiC; (b) simulated angular variation of the basal NV centers in 4H-SiC}
\label{fig:EPRdata}
\end{figure}

Here, we also report new EPR spectra of the basal NV centers in 4H SiC which have been measured in the temperature range from $4$~K to $300$~K. 
To allow comparison with the calculated values, we show in Table~\ref{tab:ZFStable3C4Ha} their parameters at T=$4$~K. The NV centers in the 4H polytype were created as reported in Refs.~\onlinecite{4Hexp, NVSiCPL,NVSiCour} by particle irradiation and thermal annealing. The observed EPR spectra are shown in Fig.~\ref{fig:EPRdata}. From these measurements, the $D$ and $E$ parameters for the basal NV configurations were extracted, and the low temperature $D$ parameters for all the four NV configurations in 4H SiC are shown in Table~\ref{tab:ZFStable3C4Ha}. We note that the previously published EPR data for axial configurations in 4H SiC are those measured at  T=300K in Ref.~\onlinecite{NVSiCour}.

\begin{table}
\begin{ruledtabular}
\caption {Ground state zero-field-splitting constants ($D$, $E$) of NV center configurations (conf.) in 3C and 4H polytypes of SiC. The experimental data ($D_\text{e}$) in 3C was taken from Ref.~\onlinecite{NVSiCour}. The experimental data in 4H SiC recorded at cryogenic temperature are also provided as new results ($D_\text{e}$, $E_\text{e}$).  In the ground state the two unpaired electrons are localized on the C atoms near Si-vacancy, hence, corresponding distances ($d_1$) are also given. For the off-axis configurations, two different C-C distances occur where the second distance is listed in the $d_2$ column. The larger the deviation is between $d_1$ and $d_2$ the larger is the $E$ parameter.}
\label{tab:ZFStable3C4Ha}
\begin{tabular}{@{}ccccccc@{}} 
conf. & $D$(MHz)& $E$(MHz) & $d_1$(\AA)& $d_2$(\AA) & $D_{\text{e}}$(MHz) &  $E_{\text{e}}$(MHz)\\ 
\hline
3C-$kk$ & 1409 & 0 & 3.34 &  & 1303 & \\
\hline
 4H-$kk$  & 1377 & 0 & 3.36 & & 1282 &  0\\
 4H-$hh$ & 1427 & 0 & 3.33 & & 1331  & 0 \\
 4H-$hk$  & 1331 & 110 & 3.38& 3.35 & 1193 & 104 \\
 4H-$kh$  & 1404 & 44  & 3.34&  3.33 &  1328 & 15 \\
\end{tabular}
\end{ruledtabular}
\end{table}
\begin{table}
\begin{ruledtabular}
\caption {Calculated values of $D$ and $E$ parameters for NV center configurations (conf.) in 6H SiC. Experimental values ($D_\text{e}$) are only available for the axial configurations\cite{NVSiCour}. Distances ($d_1$) of C-C atoms near Si-vacancy are also listed. For the off-axis configurations, two different C-C distances occur where the second distance is listed in the $d_2$ column.}
\label{tab:ZFStable6H}
\begin{tabular}{@{}ccccccc@{}}
\multicolumn{1}{c} {conf.}
& \multicolumn{1}{c} {$D$(MHz)}
& \multicolumn{1}{c} {$E$(MHz)}
& \multicolumn{1}{c} {$d_1$(\AA)}
& \multicolumn{1}{c} {$d_2$(\AA)}
& \multicolumn{1}{c} {$D_\text{e}$(MHz)}\\
\hline
$hh$  & 1404 & 0 & 3.34 & & 1328 \\
$k_1k_1$  & 1348 & 0 & 3.36 & & 1278 \\
$k_2k_2$  & 1432 & 0 &  3.33 & & 1345 \\
$k_1k_2$  & 1404  & 145 & 3.35 & 3.33 & - \\
$k_2h$  & 1386 & 9 & 3.33 & 3.34 & -  \\
$hk_1$  & 1352 & 14 & 3.34 & 3.35 & -  \\
\end{tabular}
\end{ruledtabular}
\end{table}

We showed above that proximate nuclear spins may reside around the electron spin of NV centers. The transfer of electron spin polarization to neighboring nuclear spins is principally feasible for NV center in 4H SiC and thus allows the realization of quantum memories. One possible method to spin-polarize the proximate nuclear spins is the optical dynamic polarization via excited state level anticrossing (ESLAC)\cite{JacquesPRL2009,nucspinpol}. We calculated the ZFS parameters of the $^3E$ excited states from which the required magnetic fields for ESLAC \cite{Ivady2015} can be obtained  (c.f.\ Table~\ref{tab:ZFStable3C4Hb}). 
We emphasize that ZFS parameters in the excited state are subject to dynamic Jahn-Teller effect and can be temperature dependent, and our values are only valid at $T=0$~K. In the excited state, the spin density is partially localized on the N-atom, hence, the distance between one of the C-atoms and the N-atom is given.  
\begin{table}
\begin{ruledtabular}
\caption {Calculated excited state $^3E$ zero-field-splitting constants (\textit{D}, \textit{E}) of NV center configurations (conf.) in 3C and 4H SiC. Distances between N and C atoms (N-C) around Si-vacancy are also provided.}
\label{tab:ZFStable3C4Hb}
\begin{tabular}{@{}ccccc@{}}
\multirow{1}{*} {Polytype}
& \multirow{1}{*} {conf.}
& $D$(MHz) & $E$(MHz) & N-C(\AA)\\ 
\hline
3C & $kk$ & 707.3 & 0 & 3.42  \\
\hline
\multirow{4}{*} {4H} & $kk$  & 483.0 & 0 & 3.46 \\
& $hh$ & 537.2 & 0 & 3.44  \\
& $hk$  & 471.9  & 47.8 & 3.49 \\
& $kh$  & 537.9  & 28.6 & 3.44\\
\end{tabular}
\end{ruledtabular}
\end{table}

The intracenter optical transition of the NV centers is a key parameter for all quantum applications. The zero-phonon-line (ZPL) energies have been recently identified experimentally for the four different NV centers in 4H-SiC\cite{NVSiCPL}. We calculated the ZPL energies of NV centers in SiC that is associated with the energy difference between the $^3E$ excited state and $^3A_2$ groundstate.
The results are shown in Table~\ref{tab:ZPLtable3C4H}, and once again we find a good agreement between calculated and experimental values. The discrepancy between the calculated and measured ZPL energies for each configuration is within 0.1 eV that is expected from HSE06 hybrid density functional method \cite{DeakPRB2010}. However, the calculated differences between the ZPL energies of the defect configurations are technically converged within few meV, in terms of the parameters of plane wave supercell DFT method. We found that 576-atom supercell with $2\times2\times2$ K-point sampling of Brillouin-zone is at least required, in order to obtain the correct order of ZPL energies of the four defect configurations in 4H SiC.

The accuracy of the calculated  ZPL energies is estimated to $0.1$~eV. The accuracy is limited by the supercell methodology in technical terms and by the Born-Oppenheimer approximation in the $^3E$ excited state.

Optically induced groundstate spin polarization has been observed by EPR for all NV centers in the three polytypes \cite{4Hexp, NVSiCour}. The ZPL energies correspond to the low energy threshold of the optically induced groundstate spin polarization. This feature is a particularity of all NV centers in SiC and in diamond and allows the initialization of the groundstate spin configuration by an optical pulse. It is related to the existence of intermediate singlet states, which modify the recombination processes between the $^3E$ and $^3A_2$ states. The calculation of these highly-correlated multideterminant singlet states is out of the scope of this paper. 
\begin{table}
\begin{ruledtabular}
\caption {Zero-phonon-line (ZPL) energies of individual NV center configurations in 3C and 4H SiC. Calculated and experimental ZPL data (see Ref.~\onlinecite{NVSiCPL}) are also presented.}
\label{tab:ZPLtable3C4H}
\begin{tabular}{@{}ccccc@{}}
\multirow{1}{*} {Polytype}
& \multirow{1}{*} {conf.}
& \multirow{1}{*} {ZPL(eV)}
& {Signal}
& \multirow{1}{*} {ZPL$_\text{exp}$(eV)}\\
\hline
3C & $kk$ & 0.87 & - & - \\
\hline
\multirow{4}{*} {4H} & $hh$ & 0.966 & PLX1 & 0.998\\
& $kk$  &  1.018 & PLX2 & 0.999\\
& $hk$  &  1.039 & PLX3 & 1.014\\
& $kh$  &  1.056 & PLX4 & 1.051\\
\end{tabular}
\end{ruledtabular}
\end{table}

\section{\label{sec:discussion}Discussion}

In the following sections we discuss our results  of the formation of NV centers in SiC (Sec.~\ref{subsec:discform}), the identification of individual NV centers in hexagonal SiC (Sec.~\ref{subsec:discident}), and the photo-ionization of NV centers in 4H SiC (Sec.~\ref{subsec:discphoto}). The results are  compared  to the properties of the closely related divacancy centers. Finally, we evaluate the stability of NV centers in the different SiC polytypes in Sec.~\ref{subsec:polytypes}.

\subsection{\label{subsec:discform}Formation of NV center in SiC}

We consider the formation of NV centers in 4H SiC in two scenarios: (i) formation of NV centers as native defects related to growth conditions, (ii) NV centers formed by thermal diffusion of radiation induced Si vacancies in nitrogen doped material. We neglect the kinetic effects on the surface in this study which might play a role in the first case.

\begin{figure*} [t]
\includegraphics[width=\textwidth]{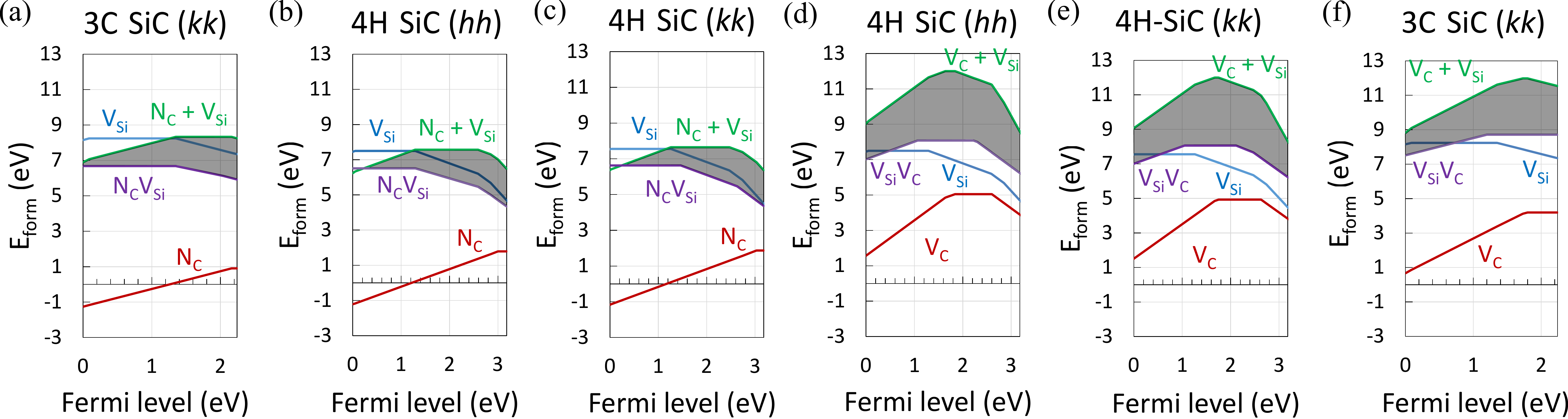}
\caption{Formation energies of axial (a)-(c) NV centers and (d)-(f) divacancies as a function of the Fermi-level in 3C and 4H polytypes. Shaded areas represent the stability of the corresponding complexes.}
\label{fig:ctlbind}
\end{figure*}

\begin{figure*} [t]
\includegraphics[width=\textwidth]{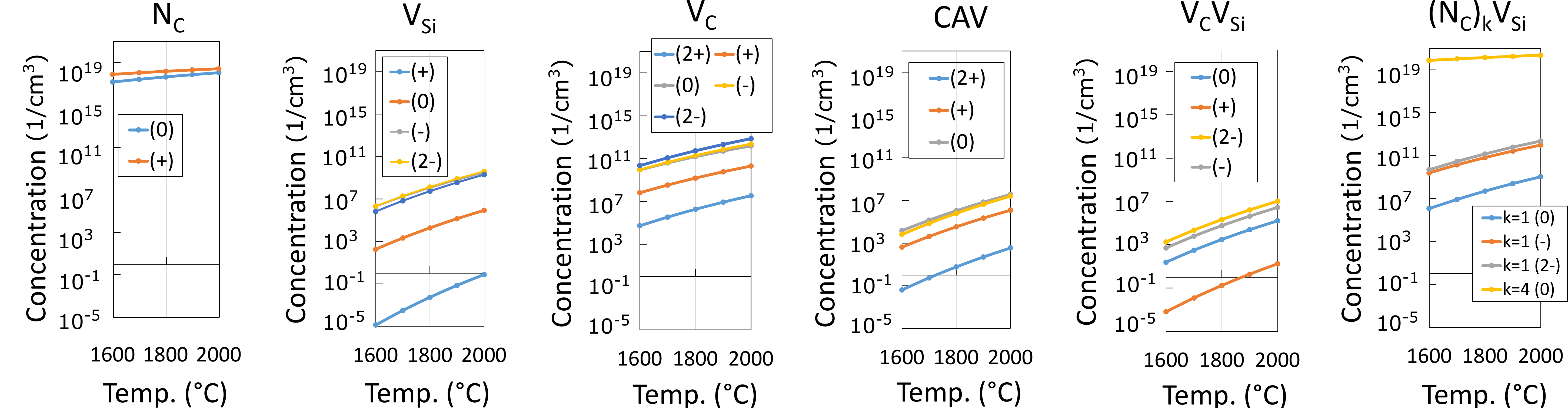}
\caption{Concentration of technologically important defects in 4H-SiC as a function of growth temperature. All the relevant charge states and configurations of each defect are considered. The concentration was calculated in thermal equilibrium assuming stochiometric SiC. Please, note the logarithmic scale on the concentration. }
\label{fig:defconc}
\end{figure*}

Homo- and hetero-epitaxial high quality thin films of SiC are generally grown by a chemical vapor deposition (CVD) process. If one approximates this CVD process as a thermal equilibrium process then the concentration of the in-grown defects can be estimated from their formation energies (see Fig.~\ref{fig:ctlbind}).  
We considered the formation of NV center by nitrogen doping of 4H SiC. We assumed the incorporation of basic intrinsic defects, C-vacancy, Si-vacancy, CAV complex, divacancy, as well as the substitutional nitrogen (N$_\text{C}$) and (N$_\text{C}$)$_4$V$_\text{Si}$ complexes beside NV center, alas, N$_\text{C}$V$_\text{Si}$ complexes. The simulations were carried out at different growth temperatures between 1600$^\circ$C and 2000$^\circ$C, typical to CVD growth of SiC. The results are plotted in Fig.~\ref{fig:defconc}.

It follows, that the concentration of in-grown Si-vacancies, CAV complexes and divacancies is negligible. (N$_\text{C}$)$_4$V$_\text{Si}$ complexes have the lowest formation energy of $\sim$1~eV and exhibit the highest concentration (over $10^{19} 1/\text{cm}^3$) which explains the known doping limitation of nitrogen in SiC (see Ref.~\onlinecite{bockstedte1}). The (N$_\text{C}$)$_4$V$_\text{Si}$ defect is electrically inactive with exhibiting $S = 0$ spin state, thus it does not establish undesirable electron spin-bath for qubit application of N$_\text{C}$V$_\text{Si}$. The maximum concentration of neutral N$_\text{C}$ is around $10^{18} 1/\text{cm}^3$. The neutral N$_\text{C}$ has an $S=1/2$ spin. The C-vacancy has a higher concentration than that of NV complexes. Nevertheless, C-vacancies will be double negatively charged at these doping conditions with a diamagnetic groundstate. 
It is important to note that the concentration of NV complex is about 7-9 orders of magnitude smaller than that of the N$_\text{C}$. In this condition, the majority of NV complex will be in the double negative charge state with $S=1/2$ spin and not in the desired single negative charge state. Whereas such low defect concentrations can be detected by PL spectroscopy, they are well below the detection limit of EPR spectroscopy. At lower temperature growth ($<1600^\circ$C) the total concentration of NV center will further decrease and be in the region where NV centers will occur as single NV centers. They can be detected by confocal PL microscopy but they have not yet been reported. However, the large concentration of neutral N$_\text{C}$ introduces a dense electron spin bath that is detrimental for single NV center ODMR measurements, because this can significantly reduce the electron spin coherence time.  

\begin{figure}
\includegraphics[width=0.450\textwidth]{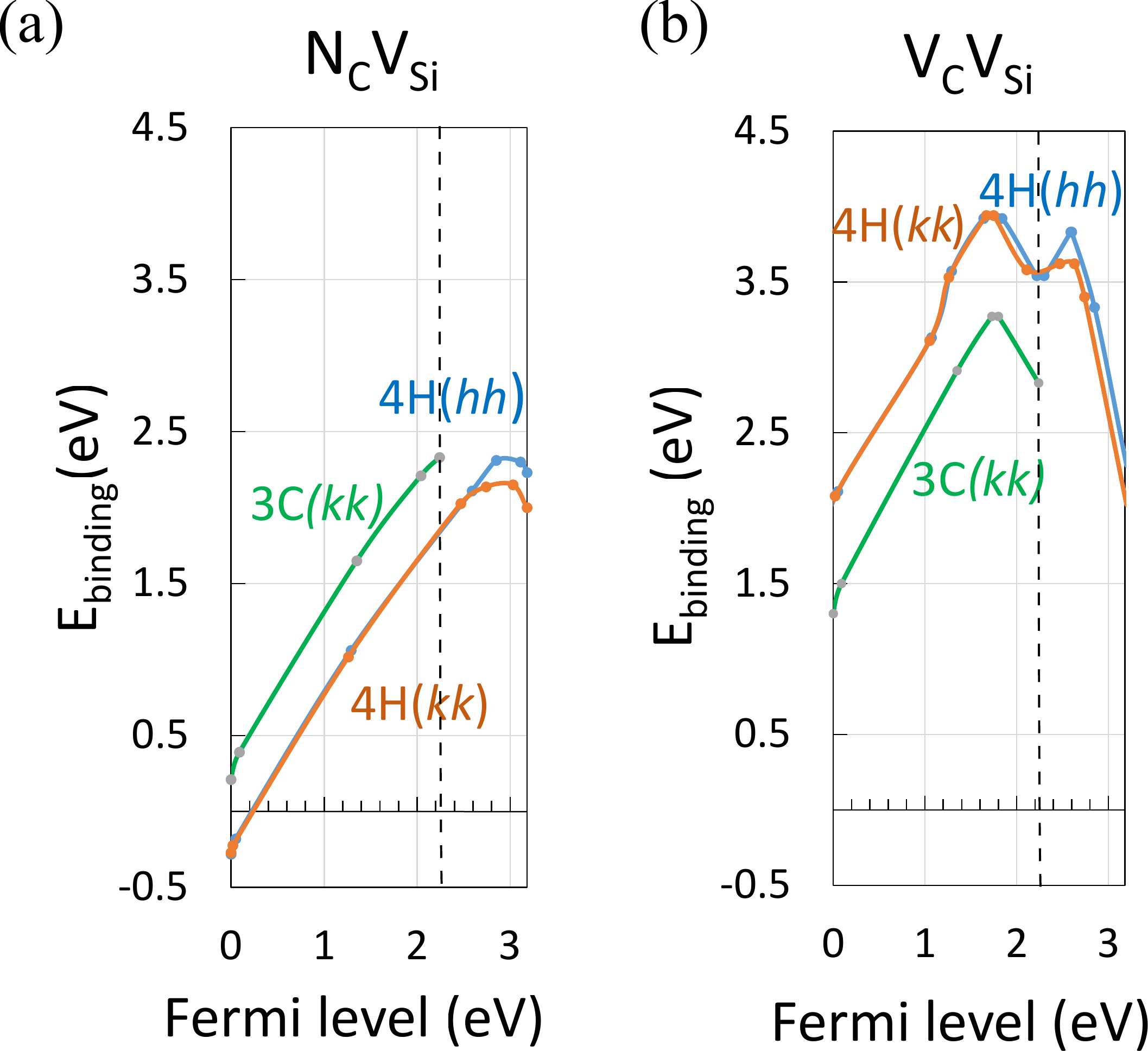}
\caption{Binding energies as a function of Fermi-level for axial (a) NV center and (b) divacancy configurations in 3C and 4H SiC. Dashed lines represent the conduction band minimum of 3C polytype that lies lower than that of 4H SiC.}
\label{fig:binde}
\end{figure}

The second approach, the one which has been successfully used in the past, is to start with lightly (10$^{16}$/cm$^3$) nitrogen doped SiC samples; then by ion implantation or particle irradiation Si-vacancies can be created;  NV center formation is obtained by thermal annealing in a temperature range where Si vacancies become mobile.  The first experimental results of  NV centers in 4H-SiC  were obtained on proton irradiated N-doped  samples \cite{4Hexp, NVSiCour}. We note that irradiation or implantation creates vacancies and interstitials in both sublattices. Regarding the monovacancies, it is expected that due to the lower displacement energies of C atoms than that of Si atoms \cite{Choyke1977, SteedsMSF2001, SteedsPRB2002, LucasPRB2005}, more C-vacancies than Si-vacancies are created by irradiation or implantation. Thus divacancies can also be be formed when Si-vacancies become mobile during annealing at around $750 ^\circ\text{C}$ \cite{Sorman2000PRB, BockstedtePRB2003}. In order to study the formation of NV centers, we calculated the binding energies of the N$_\text{C}$ and Si-vacancy versus that of the C-vacancy and Si-vacancy (see Figs.~\ref{fig:ctlbind}(a)-(c) and the derived plots in Fig.~\ref{fig:binde}). Our results show that the binding energy for divacancies is always higher  than for NV centers; consequently the formation of divacancies has a higher probability provided that C-vacancies are available for the mobile Si-vacancies. If the initial concentration of nitrogen is higher than that of C-vacancies  then the relative concentration of divacancies and NV complexes may be tuned toward the preferential formation of NV centers. This  scenario is the one used for the formation of large ensembles of NV centers. Indeed, the previous EPR and PL investigations always showed the presence of divacancies beside NV centers. \cite{4Hexp,NVSiCour}. Since divacancies have a non-zero spin in their various charge states and are optically active,  they can influence the electron spin coherence time and the photo-stability of the NV center ZPL lines. The latter will be discussed bel
 ow.  
Clearly for the formation of individual NV centers different nanoscopic approaches have to be applied, like low dose implantation  of $N_2$ ions into non-doped SiC.   

\subsection{\label{subsec:discident}Spectroscopy of individual NV centers in hexagonal SiC}

The magneto-optical parameters of large ensemble of NV centers in 4H SiC are known from EPR and PL studies \cite{4Hexp, NVSiCPL, NVSiCour}, however, future quantum technology applications will be based on the spectroscopy of individual NV centers. Therefore, it is of high importance to identify the individual NV centers in 4H SiC.   

We study first the magnetic properties. As the spin density associated with hyperfine couplings, so the spin density matrix associated with ZFS, is mostly localized on the carbon dangling bonds of the Si-vacancy in the ground state of the NV center (see Fig.~\ref{fig:spindens}). The experimentally determined\cite{4Hexp, NVSiCour} $g$-tensor anisotropy implies that the second-order spin-orbit coupling may be not negligible that may contribute to the ZFS.  Furthermore, the first order spin-orbit coupling might contribute to the ZFS in the off-axis configurations with C$_{1h}$ symmetry. Nevertheless, the vast majority of ZFS involves the electron spin - electron spin dipolar interaction, that we can fairly well calculate within Kohn-Sham DFT.

First, we discuss the on-axis configurations. The distance between carbon dangling bonds is shorter at $hh$ site than that at $kk$ site (see Table~\ref{tab:ZFStable3C4Ha}), and correspondingly, the $D$-constant is larger at $hh$ site than that at $kk$ site.  Indeed, we find the calculated $D$-constant at $hh$ configuration to be the highest one. This is very similar to the case of the divacancy in 4H SiC \cite{Falk2014}. Regarding the off-axis configurations, the $E$ constant is a good parameter to distinguish the N$_\text{C}$V$_\text{Si}$ $kh$ and $hk$ configurations. The $kh$ configuration has always a smaller $E$ constant than that of $hk$ configuration (see Table~\ref{tab:ZFStable3C4Ha}). The $hh$ and $kk$ sites can be also distinguished by their strong $^{13}$C and $^{29}$Si hyperfine interactions. Basically, the spin density is more localized in the $hh$ configuration than that in $kk$ configurations. Accordingly, the corresponding $^{13}$C hyperfine constants are higher whereas the $^{29}$Si hyperfine constants are smaller in $hh$ configuration than those in $kk$ configuration. In summary, the groundstate parameters of NV centers in 4H SiC allow a simple distinction between  the four configurations. The $\approx 100$~MHz systematic discrepancy between the calculated and experimentally determined $D$-constants might be partially attributed to the neglect of second-order spin-orbit interaction.

Another important fingerprint of the NV centers is their photoluminescence spectrum. At low temperatures, the ZPL energies in the PL spectrum of NV centers have been identified in 4H SiC \cite{NVSiCPL}. The calculated absolute values of the ZPL energies are close to the experimental ones (see Table~\ref{tab:ZPLtable3C4H}). However, the calculated site dependence of the ZPL energies is not so well reproduced by the calculations. The ZPL energy difference between the PLX1 and PLX2 NV centers is only $1$~meV. It is extremely challenging to reach such an accuracy in the calculation. We find that the $hh$ and $kk$ NV center configurations have the lowest ZPL energies, as was previously suggested \cite{4Hexp,NVSiCour}. Thus, PLX1 and PLX2 should be associated with the axial configurations. Our calculations imply that the $hh$ configuration has the lowest ZPL energy, nevertheless, the identification of PLX1 does not stand on a solid ground based on solely the calculated ZPL. In analogy with
  the divacancy, for which ODMR measurements under resonant excitation have been performed, $hh$ configuration should have the lowest ZPL energy and the largest $D$ constant \cite{Falk2014}. By assuming the same trends for the NV center, we deduce that the $hh$ configuration should be associated with the PLX1 spectrum and $kk$   with the PLX2 spectrumr. The PLX3 and PLX4 ZPL should be associated  with the off-axis NV center configurations. Our calculations imply that N$_\text{C}$V$_\text{Si}$ $hk$ and $kh$ configurations can be associated with PLX3 and PLX4 PL centers, respectively.

We also show the calculated ZFS parameters for the ground state of NV centers in 6H SiC (see Table~\ref{tab:ZFStable6H}). Interestingly, the $k_2k_2$ configuration has the largest $D$-constant which is followed by that of $hh$ and $k_1k_1$ configurations in descending order. This can be directly compared to the experimental assignments \cite{NVSiCour}. In the 6H polytype the $k_1k_2$ off-axial configuration has the largest $E$ value which is followed by that of $hk_1$ and $k_2h$ configurations in descending order. The PL spectrum of the NV centers in 6H SiC has not yet been reported.

\subsection{\label{subsec:discphoto}Photo-ionization of NV centers in 4H SiC}

The photo and spectral stability of solid state qubits is of high importance. The photostability of NV centers might be compromised by simultaneous photo-ionization. The   photon ionization energies are given in Table~\ref{CTLtab}: the NV center in a negative charge state NV($-$) can be photo-ionized to neutral charge state NV($0$) by promoting an electron to the conduction band edge with an energy of about 1.7-1.8~eV. This energy is sufficient to re-ionize NV($0$) to NV($-$) by promoting an electron from the valence band to the in-gap defect level. In the case of single defect spectroscopy, confocal microscopy is applied with high excitation power that can result in two-photon absorption processes. According to our results,  two-photon absorption process will also give rise to photoionisation. Unfortunately, the excitation energy of NV($0$) is not known and cannot be calculated by Kohn-Sham DFT. Therefore, the threshold energy to reionize NV($0$) to NV($-$) by two-photon absorption 
 is  not known. If the ZPL energy of NV($0$) is higher than that of NV($-$) then NV($-$) should be excited into the phonon sideband, in order to re-ionize NV($0$) to NV($-$) by two-photon absorption.  

Spectral diffusion in the emission of single NV($-$) center might occur upon photo-excitation when nearby defects are simultaneously excited resulting in fluctuating charges in the proximity of the NV($-$) center. These fluctuating charges can shift the ZPL energy of NV($-$) by small amounts which is detrimental for  quantum applications. We showed above that divacancy defect will form together with NV centers under implantation or irradiation induced formation. Our calculations show, if divacancies are excited by about 1.1~eV  two-photon absorption   can also occur which leads to the ionization of the neutral divacancy [see Fig.~\ref{CTLfig}(c)]. The same photon energy is sufficient to excite NV($-$) in the phonon side band. Our calculations imply that excitation energy lower than 1.1eV should be applied, in order to avoid the photo-ionization of the divacancies.

We may conclude by emphasizing the two conditions for optimal readout process of  NV qubits in 4H SiC, (i) efficient re-ionization of NV($-$) and (ii) conserving spectral stability by avoiding ionization of the divacancy by two-photon absorption. 

\subsection{\label{subsec:polytypes}Comparison of SiC polytypes to host NV qubits}

We previously discussed the photo and spectral stability of NV qubits in 4H SiC in detail. Here, we further discuss this issue for NV qubits in 3C and 6H SiC.

In 3C SiC, the calculated acceptor level of NV defect lies at about 1.5~eV with respect to valence band edge, similarly to the NV defects in the hexagonal polytypes. Since the band gap of 3C SiC is smaller, 2.4~eV at low temperatures, the ZPL energy for the intra center transition and the ionization energy of NV($-$) almost coincide.
Thus excitation of NV center in 3C-SiC results in an excited state resonant with the conduction band edge. In this case the photostability of NV($-$) in 3C SiC may be difficult to maintain, particularly, at elevated temperatures. We note that the ensembles of NV centers could be efficiently optically spinpolarized in N-doped 3C SiC \cite{NVSiCour}. Nevertheless, this process might involve recapture of free electrons created by photoexcitation, that may not efficiently work at single defect level. On the other hand, if a spin-selective photo-ionization occurs for this defect then photo-ionization based detection of magnetic resonance (PDMR) could be the appropriate methodology \cite{PDMR} to read out the spin of single NV qubit.  

The band gap of 6H SiC is about 0.25~eV lower than that of 4H SiC. The acceptor levels with respect to the valence band edges and excitation energies of the NV defects are very similar in the two polytypes. As a consequence, the two-photon absorption of NV($-$) defect in 6H SiC may be more effective than that in 4H SiC because of the larger density states in the conduction bands of 6H SiC. Thus, the relative rates of ionization and re-ionization of NV($-$) centers in 6H SiC may shift toward the ionization and  compromise the stability of NV centers in 6H SiC. If NV($-$) is excited with an energy above the ZPL energy of divacancy then two-photon ionization of neutral divacancy can take place because of the relatively low-lying conduction band edge of 6H SiC. We conclude that our calculations imply that 4H SiC with the largest band gap is the optimal host for NV center  qubit applications  with optical read out.

\section{\label{sec:summary}Summary}

In summary, we carried out DFT calculations of N$_{\text{C}}$V$_{\text{Si}}$ defects in 3C, 4H and 6H SiC. We focused on the negatively charged N$_{\text{C}}$V$_{\text{Si}}$ defect, i.e., the NV center with potential qubit applications. We
discussed the formation of NV center in SiC, and found the divacancy inevitably forms when NV center is produced by implantation or irradiation. We calculated the ground state and excited state magneto-optical properties of the NV centers and compared them with the experimental magneto-optical data. We identified the individual NV qubits in hexagonal polytypes. We also discussed the photo-ionization and spectral stability of NV center in SiC. Our results show the importance of selective photoexcitation avoiding simultaneous excitation of divacancies in the two-photon absorption process.

\section*{\label{sec:ackn} Acknowledgments}
Andr\'as Cs\'or\'e acknowledges the Ministry of Human Resources for financial support in the framework of National Talent Program (NTP-NFT\"O-16). Adam Gali acknowledges Hungarian NKFIH grant No.~NVKP\_16-1-2016-0152958.


%

\end{document}